\documentclass[twocolumn,showpacs,preprintnumbers,amsmath,amssymb]{revtex4}

\usepackage{graphicx}
\usepackage{epstopdf}
\usepackage{dcolumn}
\usepackage{bm}

\usepackage{amsmath}
\usepackage{amsfonts}
\usepackage{amssymb}

\begin{document}


\title{The Strategy of Discrimination between Flavors for Detection of Cosmogenic Neutrinos}

\author{Kwang-Chang Lai}
 \email{kcl@mail.cgu.edu.tw}
\affiliation{Center for General Education, Chang Gung University, Kwei-Shan, Taoyuan, 333, Taiwan\\
Leung Center for Cosmology and Particle Astrophysics (LeCosPA), National Taiwan University, Taipei,106, Taiwan}

\author{Chih-Ching Chen}
 \email{b88202054@ntu.edu.tw}
\affiliation{Graduate Institute of Astrophysics, National Taiwan University, Taipei,106, Taiwan.\\
Leung Center for Cosmology and Particle Astrophysics (LeCosPA), Taipei, 106, Taiwan.}

\author{Pisin Chen}
 \email{chen@slac.stanford.edu}
\affiliation{Graduate Institute of Astrophysics, National Taiwan University, Taipei,106, Taiwan.\\
Department of Physics, National Taiwan University, Taipei, 106, Taiwan.\\
Leung Center for Cosmology and Particle Astrophysics (LeCosPA), Taipei, 106, Taiwan.\\
Kavli Institute for Particle Astrophysics and Cosmology, SLAC National Accelerator Laboratory, Menlo Park, CA 94025, USA}

\date{\today}

\begin{abstract}

We propose a new method to identify flavors of ultra high energy cosmic neutrinos. Energy loss of leptons in matter provides important informations for the detection of neutrinos originated from high energy astrophysical sources. 50 years ago, Askaryan proposed to detect Cherenkov signals by radio wave from the negative charge excess of particle showers. The theory of Cherenkov pulses with Fraunhofer approximation was widely studied in the past two decades. However, at high energies or for high density materials, electromagnetic shower should be elongated due to the Landau-Pomeranchuck-Migdal (LPM) effect. As such the standard Fraunhofer approximation ceases to be valid when the distance between the shower and the detector becomes comparable with the shower length. We have performed Monte Carlo simulations recently to investigate this regime based on the finite-difference time-domain (FDTD) method, and modified time domain integration method. In this work, we adopt the deduced relationship between the radio signal and the cascade development profile to investigate its implication to lepton signatures. Our method provides a straightforward technique to identify the neutrino flavor through the detected Cherenkov signals.

\end{abstract}

\pacs{Valid PACS appear here}
\maketitle

\section{Introduction}

The nature and origin of ultra-high energy cosmic rays (UHECRs) have remained a mystery. These amazingly energetic events have been observed beyond $\approx$ $10^{19.6}$ eV, the so-called Greisen-Zatsepin-Kuzmin(GZK)~\cite{gzk_process} cut-off. The GZK feature on the UHECR spectrum has been first observed by the High Resolution Fly's Eye Experiment~\cite{cutoffHiRes} and later confirmed by the Pierre Auger Observatory~\cite{cutoff}. Above this energy scale, UHECRs interact with CMB photons through the GZK processes~\cite{gzk_process}, producing cosmogenic neutrinos. The GZK feature on the cosmic ray energy spectrum guarantees the existence of the cosmogenic neutrinos. However, none of these have been observed so far. Detecting these ultra high energy (UHE) neutrinos provides critical informations for unraveling the mystery of the origin and evolution of the cosmic accelerators and will be one of the utmost tasks in the coming decade~\cite{pisin_whitepaper}.

One promising way of detecting UHE neutrinos is the radio approach. When an ultra-high energy cosmic neutrino interacts with ordinary matters on the Earth, it would lead to a hadronic debris, either by charged current or neutral current. The former also produces a lepton with corresponding flavor. Both the high energy leptons and the hadronic debris induce particle showers. As proposed by Askaryan in the 1960's~\cite{Askaryan}, the high energy particle shower develops in a dense medium would have net negative charges. This charge imbalance appears as a result of the knocked-off electrons being part of the shower, as well as the positrons in the shower annihilating with the electrons of the medium. The net charges of the showers, typically $20 \%$ of total shower particles, serve as a source emitting the Cherenkov radiations when they travel in the medium. The sizes of the showers are quite localized (tens of cm in radial and few meters in longitudinal development) compared to those develop in the air (km scale), and therefore result in coherent radiations for wavelengths longer than the shower sizes. The corresponding coherent wavelength turns out to be in the radio band, from hundreds of MHz to few GHz.

In this article, we discuss the possibility to identify the flavors of the cosmogenic neutrinos detected by the radio neutrino telescope, such as ANITA~\cite{ANITA},  Askaryan Radio Array (ARA)~\cite{ARA}, and ARIANA~\cite{ARIANA}.

\section{The Strategy of Flavor Identification}

As neutrinos interact with matters to produce observable signals, the major channel is the changed-current (CC) interaction. The electron produced through $\nu_e$ CC interaction has a large interaction cross section with the medium and produces a shower within a short distance from its production point. Contrary to the electron, the muon produced through $\nu_\mu$ CC interaction can travel a long distance in the medium before it loses all its energy or decays. However, a muon does emit dim light along its propagation so that only those detectors near to the muon track can be triggered. 

As for $\nu_\tau$ detection, the $\nu_\tau$-induced tau leptons behave differently at different energies for a fixed detector design. For a neutrino telescope such as IceCube, the observable energy range for the double bang event is $3.3{\rm PeV}<E_\nu<33{\rm PeV}$. For an undersea experiment, such as KM3Net~\cite{KM3Net}, the observable energy range for the double bang event is similar. But, for a radio neutrino telescope, such as ARA, the detector is designed to observe cosmogenic neutrinos of energy about EeV. In this energy regime, the tau lepton range becomes long enough so that a tau lepton can pass through the detector without decaying but losing its energy like a muon does. In this case, the signal for $\nu_\tau$ appears like a track event.

Note that the dim lights emitted from the track can only trigger the nearby optical detectors but cannot be received by radio detectors. A different strategy is taken to construct track events for radio detectors. For cosmogenic neutrinos, the energy of the CC-induced muon or tau lepton is so high that a muon or tau lepton not only emits dim lights but also produce mini-showers along its propagation through the detector fiducial volume. By detecting the radio emissions from these mini-showers, a track event is reconstructed for a muon or tau lepton traveling through the detector. By observing a single shower, a $\nu_e$ signal is identified from a track event for a $\nu_\mu$ or $\nu_\tau$.

It is challenging to distinguish between $\nu_\mu$ and $\nu_\tau$ signals because both muons and tau leptons produce similar track-like events. Simulation of lepton propagation in ice shows that the compositions of the mini-showers are different for muon and tau lepton track events. The mini-showers that consist of the track events are composed of two categories, electromagnetic (EM) and hadronic showers. The energy loss distribution between EM and hadronic showers is different for muon and tau lepton track events. A muon track event loses more energy through EM showers than through hadronic ones while a tau track loses more energy through hadronic showers than through EM ones. By collecting mini-showers, measuring their attributes and evaluating energy losses, one can distinguish between muon and tau track events.

\section{Simulation for Neutrino Events}

\begin{figure}[htbp]
	\begin{center}
	\includegraphics[width=8cm]{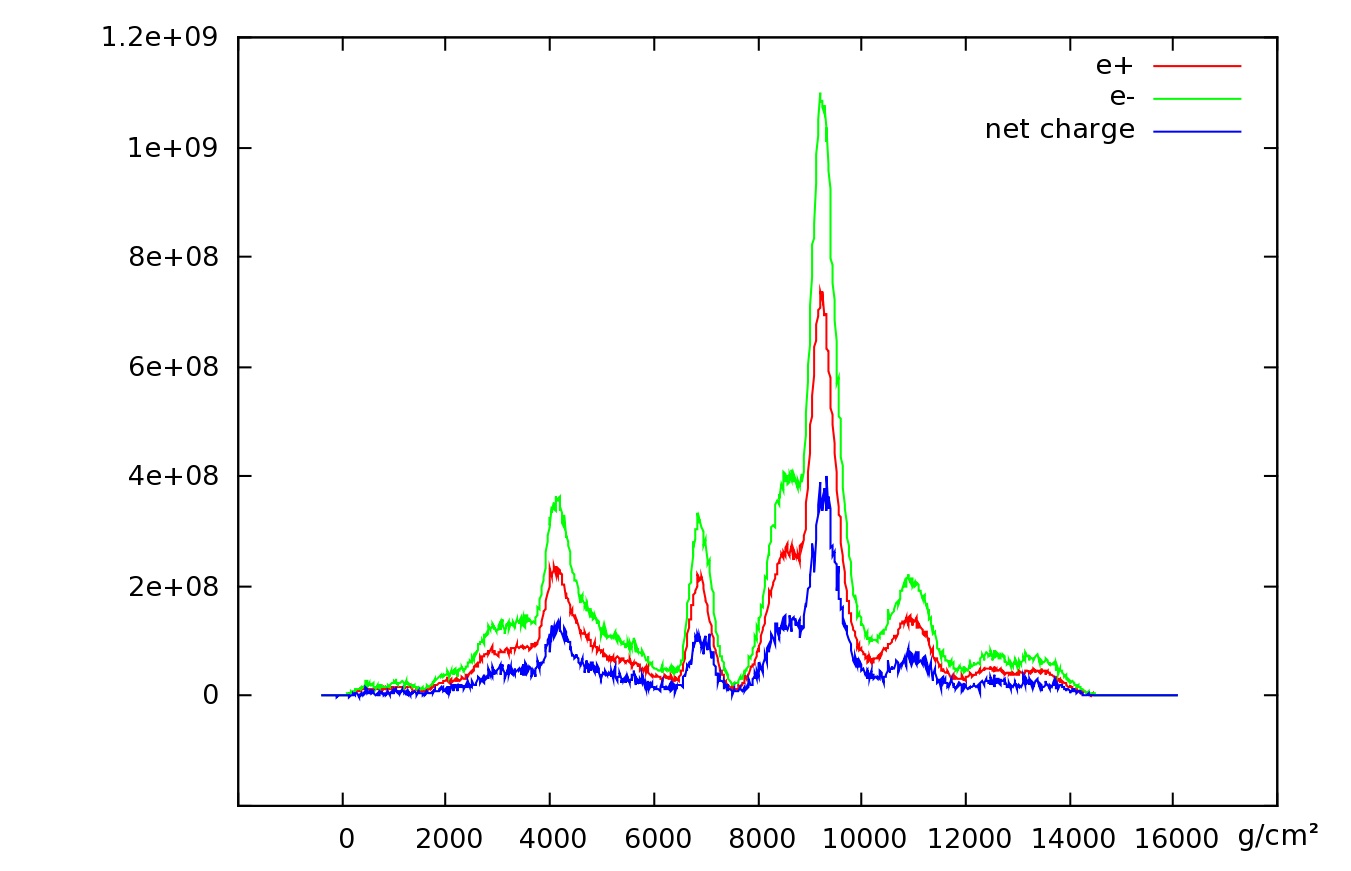}
	\caption{Longitudinal profile of a $10^{19}$ eV electron shower in ice. As LPM suppression increases with the square root of the energy, multiple peaks occur during the elongated shower development. The profile is sensitive to the initial interactions of the cascade.}
	\label{fig:ele19ev}
	\end{center}
\end{figure}

\begin{figure}[htbp]
	\begin{center}
	\includegraphics[width=8cm]{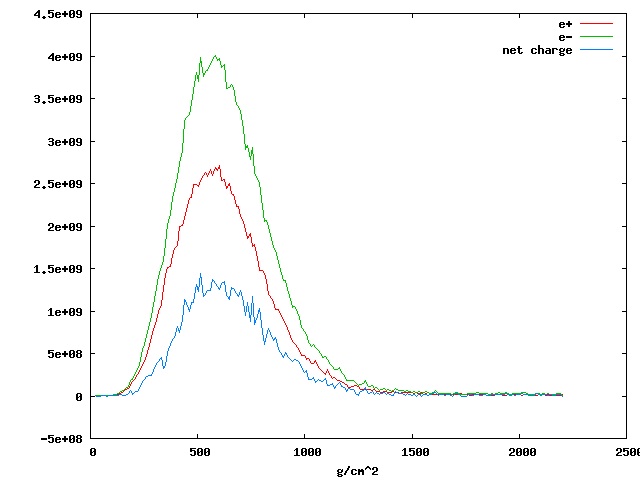}
	\caption{Longitudinal profile of a $10^{19}$ eV proton shower in ice. The hadronic shower is initiated by the cascade of mesons. The decay length of neutral pions is longer than the LPM interaction length in ice for electrons at energy $10^{15}$ eV. Meanwhile the secondary mesons produce very few electrons due to LPM suppression.}
	\label{fig:pro19ev}
	\end{center}
\end{figure}

ARA detectors receive radio emissions from shower particles created by cosmogenic neutrinos in ice. These radio signals are Cherenkov radiations produced by net charges of showers particles. We adopt COSIKA-IW~\cite{Bolmont,acorne} code, a modification of COSIKA~\cite{Heck} program for dense-target simulation, to simulate EM and hadronic showers. In Fig. \ref{fig:ele19ev} and \ref{fig:pro19ev}, longitudinal developments of charges are shown for EM and hadronic showers respectively. The hadronic shower is simulated by a proton hitting a dense medium and evolves as a typical profile with one peak at the shower maximum. But,f for the EM shower, the case is different. At energies higher than $10^{16}{\rm eV}$, bremstrahlung and pair-production processes are suppressed by Landau-Pomeranchuck-Migdal~\cite{LPM} effect. As a result, cascades are stretched, shower development is elongated and several peaks appear on the profile.

With the shower profile, Cherenkov radiations can be simulated and evaluated with the time-domain finite-difference (FDTD) method\cite{EM_FDTD}. Between the shower profile $\rho(x)$ and the radio signal $E(t)$ exists a one-on-one correspondence~\cite{BR}. The electric field of Cherenkov radiation can be calculated by solving the inhomogeneous Maxwell equations, as it has been demonstrated by Alvarez-Muniz et al~\cite{AZ_flavor}. EM and hadronic showers produce signals received in different patterns in radio detectors. By measuring radio signals, showers are identified and their energies are inferred.

\begin{figure}[htbp]
	\begin{center}
	\includegraphics[width=8cm]{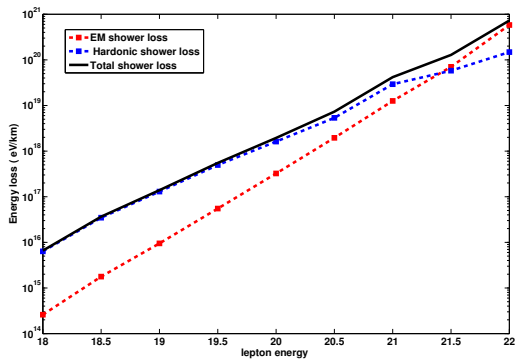}
		\caption{Energy loss for tau track. The dominant energy loss process of tau propagating in ice is photonuclear interaction. Taus energy loss rate through EM processes is suppressed since the cross-sections of pair production and bremsstrahlung are inversely proportional to the lepton mass squared.}
	\label{fig:tauloss}
	\end{center}
\end{figure}

\begin{figure}[htbp]
	\begin{center}
	\includegraphics[width=8cm]{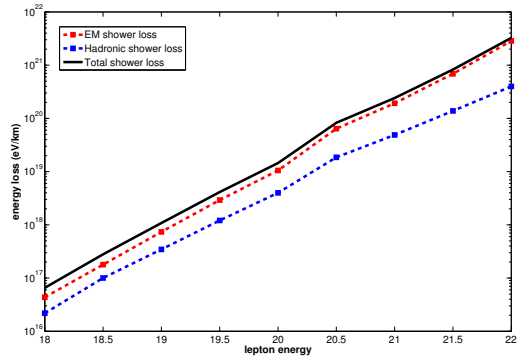}
	\caption{Energy loss for muon track. The dominant energy loss processes of muon propagating in ice are pair production and bremsstrahlung. The secondary particles (e+, e?,) from those processes
generate the electromagnetic shower. As can be seen, the EM shower profile is distinguishable
from the hadronic one.}
	\label{fig:muloss}
	\end{center}
\end{figure}

To study muon and tau tracks, we adopt the muon Monte Carlo package to simulate muon and tau lepton propagations in ice. In Fig. \ref{fig:tauloss} and \ref{fig:muloss}, energy loss from EM and hadronic showers are shown for muon and tau tracks respectively. For muon propagation, EM processes are dominant mechanism to lose energy and hadronic processes are subdominant.  For tau lepton propagation, the case is reversed. Moreover, the energy loss distribution among different types of showers depends upon track energy. Once a track is sufficiently measured, its type is identified and its energy can be inferred as well.

\section{Summary}

In this work, we propose our strategy to identify the neutrino flavor in observing cosmogenic neutrinos by radio neutrino telescopes, such as ARA. We point out that $\nu_e$ can be identified from $\nu_\mu$ and $\nu_\tau$ because $\nu_e$ produces an EM shower in ice while $\nu_\mu$ and $\nu_\tau$ produce track events. We also propose to construct these tracks by detecting those mini-showers emerged along the lepton propagation. To distinguish between $\nu_\mu$ and $\nu_\tau$, lepton propagation in ice is simulated with MMC. We find that energy loss distribution among EM and hadronic showers depends on both lepton identity and energy. Simulations for shower production with COSIKA-IW show different particle profiles for EM and hadronic showers so that the shower type and its energy can be inferred with FDTD method.

In summary, neutrino flavors can be discriminated between one another for cosmogenic neutrinos with radio neutrino telescopes. We will refine our method of flavor discrimination with more detailed study on shower production, lepton propagation and radiation conversion.

\section*{Acknowledgements}

We would like to thank Albrecht Karle, David Besson, Peter Gorham and David Seckel for valuable discussions. This research is supported by Taiwan National Science Council (NSC) under Project No. NSC-100-2119-M-002-525, No. NSC-100-2112-M-182-001-MY3 and US Department of Energy under Contract No. DE-AC03-76SF00515. We would also like to thank Leung Center for Cosmology and Particle Astrophysics for its support.

\end{document}